\documentstyle[epsf,aps,twocolumn]{revtex}
\newcommand{\be}{\begin{equation}}   \newcommand{\ee}{\end{equation}}
\newcommand{\ba}{\begin{array}}      \newcommand{\ea}{\end{array}}
\newcommand{\bea}{\begin{eqnarray}}  \newcommand{\eea}{\end{eqnarray}}
\begin{document}
\twocolumn[\hsize\textwidth\columnwidth\hsize\csname
@twocolumnfalse\endcsname

\title{Effect of Long-Range Interactions in the Conserved 
Kardar-Parisi-Zhang Equation }
\author{Youngkyun Jung and In-mook Kim\cite{mail2}}
\address{Department of Physics, Korea University, Seoul 136-701, KOREA}
\author{Jin Min Kim}
\address{Department of Physics, Soongsil University, Seoul 156-743, KOREA}
\date{\today}
\maketitle
\begin{abstract}
The conserved Kardar-Parisi-Zhang equation in the presence of long-range 
nonlinear interactions is studied by the dynamic renormalization group method.
The long-range effect produces new fixed points with continuously varying
exponents and gives distinct phase transitions, 
depending on both the long-range interaction strength and 
the substrate dimension $d$. 
The long-range interaction makes the surface width less rough than that
of the short-range interaction. In particular,
the surface becomes a smooth one with a negative roughness exponent 
at the physical dimension $d=2$
\bigskip
\draft
\end{abstract}
]

For the last decade the kinetic roughening of surfaces has 
attracted much interest\cite{dyn}.
The recent studies concentrate on measuring the scaling exponents  
which characterize the asymptotic behavior of the surface roughness on a 
large length scale and in a long time limit, 
and finding the continuum equations. 
The problem of rough surface is not only of practical 
importance in crystal growth, but also related to the nonequilibrium 
statistical physics.
Lots of computer simulations and theoretical approaches have been applied
in the studies of the Kardar-Parisi-Zhang (KPZ)\cite{kpz} equation and 
discrete molecular-beam-epitaxy (MBE) growth models 
with various kind of noises\cite{laidas,medina,amar,lam,sgg}. 
Among them, the Eden model\cite{eden}, ballistic deposition\cite{ballistic}, 
and the restricted solid-on-solid (RSOS) growth model\cite{rsos}
have been identified 
as a universality class corresponding to the KPZ equation for the coarse
grained height variable $h({\bf r},t)$ which describes the surface as a 
function of coordinate ${\bf r}$ and time $t$.
The KPZ equation has a nonlinear term of short range describing the
lateral growth. 
However, there is a poor agreement between the KPZ equation 
and the experimental data. 

Recently, Mukherji and Bhattacharjee\cite{mukherji} proposed 
a phenomenological equation in the presence of long-range interactions
to describe the kinetic roughening of the surface growth.
The long-range effect of the nonlinear term in the KPZ 
equation is introduced by coupling the gradients at two different points. 
The roughness of the surface are found to depend on the long-range nature
and several distinct phase transitions are observed.
The long-range interactions decaying slower
than $1/r^{d}$ ($d$ is the substrate dimension) makes the KPZ fixed point
with the short-range interaction be unstable.
The surface then have the long-range roughness with different exponents
depending on the power law of the long-range interactions. 
Other interactions decaying faster than $1/r^{d}$ are
suppressed by the local interaction yielding the  
ordinary KPZ universality class.

In the kinetic roughening problems, the
universality class of the dynamic systems depends on  
the symmetry of the order parameter, the dimensionality of space, 
and the conservation of the surface currents.
Therefore it would be interesting to examine how the long-range 
interaction in the conserved growth equation affects both the roughness of 
the surface and the phase transitions compared to the cases of the 
short-range interaction. We extend the  
phenomenological equation of Mukherji and Bhattacharjee, 
to a conserved equation,
\bea
\frac{\partial h({\bf r},t)}{\partial t}&=&
 - K \nabla^{4}h({\bf r},t) + \eta_{c}({\bf r},t) \nonumber \\
 -\frac{1}{2}&\nabla^{2}& \int d{\bf r}'{\vartheta}({\bf r}')
       \nabla h({\bf r+r'},t) \cdot \nabla h({\bf r-r'},t)~~,
                   \label{lrckpz}
\eea
where $h({\bf r},t)$, assumed to be a single-valued function of 
position ${\bf r}$, describes the height of the surface.
The parameter $K$ is a constant, and
$\eta_{c}$ is a conserved random noise of zero mean
with $\langle \eta_{c}({\bf r},t) \eta_{c}({\bf r}',t')\rangle
=-2D_{c} \nabla^{2}\delta({\bf r}-{\bf r}')\delta(t-t')$.
Since the right hand side of Eq.~(\ref{lrckpz}) can be
written as the divergence of a current, the total volume under the
surface is conserved.
The kernel ${\vartheta}({\bf r})$ includes long-range part which is
connected to the underlying interactions. As Ref.~ \cite{mukherji}, 
we take ${\vartheta}({\bf r})$
to have a short-range(SR) part $\lambda_{0}\delta({\bf r})$
and a long-range(LR) part $\sim r^{\rho-d}$, or more precisely,
in Fourier space, ${\vartheta}({\bf k})=\lambda_{0}+\lambda_{\rho}k^{-\rho}$.

The surface width $W(L,t)$ can be described by the dynamical scaling form,
$W(L,t)=L^{\chi}F(t/L^{z})$, where $L$, $\chi$, $z$ and $F$ are the
system size, the roughness exponent, the dynamic exponent, and 
the scaling function respectively.
For $\lambda_{0}=\lambda_{\rho}=0$, it becomes a linear equation 
evolving with the conservative surface diffusion, where the roughness 
exponent $\chi$ is $(2-d)/2$, and the dynamic exponent $z$ is four. 
For the physical dimension $d=2$, $\chi=0$, thus the surface width is  
logarithmically rough as a function of system size $L$. 
Above two dimensions the linear equation with the 
conservative noise produces negative roughness exponent, implying
a smooth surface\cite{yang}.
For $\lambda_{\rho}=0$ and $\lambda_{0} \neq 0$, Eq.~(\ref{lrckpz}) becomes 
the conserved KPZ equation with a conservative noise 
(called Sun-Guo-Grant(SGG) equation)\cite{sgg,imkim}, 
where the average height remains constant. 
For this local conserved growth equation,  
the dynamic renormalization group(RG) calculation shows 
$\chi=(2-d)/3$ and $z=(10+d)/3$\cite{sgg}.
For $d \geq 2$, the nonlinear term is irrelevant and then 
the exponents are given by the linear theory with both
$\lambda_{0}$ and $\lambda_{\rho}$ being zero in Eq.~(\ref{lrckpz}).  
Here we show that
a long-range part($\lambda_{\rho} \neq 0$) gives a new fixed point 
with continuously varying exponents and thus yields distinct phase 
transitions depending on both the parameter$\rho$ of the long-range 
interactions and the substrate dimension $d$.
This nonlocal $\lambda_{\rho}$ term with positive $\rho$ makes the surface
less rough than the case of $\lambda_{\rho}=0$.
Especially, at the physical dimension $d=2$, the surface becomes a smooth phase
with a negative roughness exponent rather than logarithmically 
rough phase as in the SGG case.

Under the change of scale, the parameters in Eq.(\ref{lrckpz}) change to 
$ K \rightarrow b^{z-4} K$, $D_{c} \rightarrow b^{z-2\chi-d-2}D_{c}$,
$\lambda_{0} \rightarrow b^{z+\chi-4} \lambda_{0}$, and
$\lambda_{\rho} \rightarrow b^{z+\chi+\rho-4} \lambda_{\rho}$.
In the absence of nonlinearity($\lambda_{0}=\lambda_{\rho}=0$), $K$
and $D_{c}$ are scale invariant to yield $z_{0}=4$ and $\chi_{0}=(2-d)/2$.
Using these values we find that the nonlinearities rescales as
$\lambda_{0} \rightarrow b^{(2-d)/2} \lambda_{0}$ and
$\lambda_{\rho} \rightarrow b^{(2+2\rho -d)} \lambda_{\rho}$.
So that the critical dimensions are given by $d_{c}=2+2\rho$ ($\rho >0$) 
and $d_{c}=2$ ($\rho <0$) for any nonzero $\lambda_{\rho}$.
When $\rho >0$, if $d<d_{c}=2+2\rho$, 
the fixed point of the local interaction ($\lambda_{\rho}=0$,
$\lambda_{0} \neq 0$ and $z+\chi-4=0$) 
is unstable and thus a new fixed point is expected.
If $2+2\rho \leq d$,  the nonlinearities become irrelevant and the surface
is controlled by the linear equation.
For $\rho<0$, if $d<2$, the SGG fixed point is stable so that 
$\lambda_{0}$ is relevant rather than $\lambda_{\rho}$, otherwise the linear
term is relevant. 
As a result, various phase diagrams depending on $d$ and $\rho$ 
would be appeared.

Following a dynamic RG procedure\cite{medina,fns}, 
integrating out fast modes 
in the momentum shell $e^{-\ell}\Lambda \leq |{\bf k}| \leq \Lambda$ and 
performing the rescaling $r \rightarrow br$, $t \rightarrow b^{z}t$, 
$h \rightarrow b^{\chi}h$,
we derive the following flow equations for the coefficients,
in a one-loop approximation,
\bea
\frac{d K}{d \ell}&=&K[z-4-\frac{D_{c} B_{d}}{K^{3}}
   {\vartheta}(1)\frac{d-4+3f(1)}{4d}]  \label{kappa}\\
\frac{d D_{c}}{d \ell}&=&D_{c}[z-2\chi-d-2] \label{delta} \\
\frac{d \lambda_{0}}{d \ell}&=&\lambda_{0}[z+\chi-4] \label{lam0} \\
\frac{d \lambda_{\rho}}{d \ell}&=&\lambda_{\rho}[z+\chi-4+\rho] \label{lamrho}
\eea
where $f(a)=\partial {\rm ln} {\vartheta}(k)/\partial {\rm ln} k|_{k=a}$, and 
$B_{d}=S_{d}/(2\pi)^{d}$, $S_{d}$ being the surface area of a 
$d$-dimensional unit sphere. 
Since the diagrams contributing to $D_{c}$ have prefactors proportional
to $k^4$, they correspond to higher derivatives in the original
noise spectrum.
Note that two scaling relations, $z+\chi =4$ and $z+\chi =4-\rho$
which result from the non-renormalization of the $\lambda_{0}$
and $\lambda_{\rho}$ in Eq.(\ref{lam0}) and Eq.(\ref{lamrho}) respectively, 
are the results of one-loop approximation\cite{janssen}.

Defining the dimensionless parameters
$U^{2}_{0} \equiv (D_{c} \lambda^{2}_{0} B_{d})/K^{3}$,
$U^{2}_{\rho} \equiv (D_{c} \lambda^{2}_{\rho} B_{d})/K^{3}$,
and $R=U_{0}/U_{\rho}$,
we obtain the flow equations for $U_{0}$, $U_{\rho}$, and $R$:
\bea
\frac{dU_{0}}{d \ell}\! &=& \! U_{0}\! \left[ \frac{2-d}{2}
                       \! +\! \frac{3(d-4)}{8d}U^{2}_{0}
 \! + \! \frac{3U_{\rho}}{8d}(c_{0}U_{0}\! +\! c_{1}U_{\rho}) 
                               \right] \label{u0} \\ 
\frac{dU_{\rho}}{d \ell}\! &=&\! U_{\rho}\! \left[\frac{2-d+2\rho}{2}
         \! +\! \frac{3(d-4)}{8d}U^{2}_{0} \right. \nonumber \\ 
& & \ \ \ \ \ \ \ \ \ \ \ \ \ \ \ \ \ \ \  
            \left. +\frac{3U_{\rho}}{8d}(c_{0}U_{0}+c_{1}U_{\rho}) 
                      \right] \label{urho}
\eea
and $dR/d\ell =-\rho R$,
where $c_{0}=(d-4)2^{-\rho}+d-4-3\rho$, and $c_{1}=(d-4-3\rho)2^{-\rho}$.  
The equation for $R$ rules out the existence of any off-axis fixed point
in the $U_{0}$ and $U_{\rho}$ parameter space (except for $\rho =0$).
From these equations we find that there are only two sets of axial fixed 
points in the two dimensional ($U_{0},U_{\rho}$) space:
SR $\equiv$ $\{U_{0}^{*2}=4d(d-2)/3(d-4), U_{\rho}^{*2}=0\}$,
with $\chi +z=4$, and LR $\equiv$ $\{U_{0}^{*2}=0,  
U_{\rho}^{*2}=4d(d-2-2\rho)/3(d-4-3\rho)2^{-\rho}\}$, with $\chi +z=4-\rho$.
When $U_{\rho}=0$, the SR fixed point is stable for $d<2$, 
where $\chi=(2-d)/3$ and $z=(10+d)/3$, 
in agreement with the results of Sun, Guo, and Grant\cite{sgg}. 
For $d \geq 2$, $U_{0}$ is driven to zero as $\ell \rightarrow \infty$.
The surface width is thus described by the linear equation yielding 
a smooth phase except for $d=2$(logarithmically rough phase).
Similarly, from Eq. (\ref{urho}) with $U_0 =0$,
the LR fixed point for $d<2+2\rho$ is stable.
At this new LR fixed point, the exponents are given by
\bea
\chi =(2-d-\rho)/3,\,\,\, {\rm and} \,\,\, 
                z=(10+d-2\rho)/3. \label{exponent} 
\eea
These exponents are determined by Eq.(\ref{delta}) and
Eq.(\ref{lamrho}) in which $D_{c}$ and $\lambda_{\rho}$ are not
renormalized in an one-loop approximation ($z-2\chi -d-2=0$ and 
$z+\chi =4-\rho$, respectively).

From these recursion relations, we can discuss the surface morphologies and 
the phase transitions for all $d$'s and $\rho$'s (see FIG.~1).
Note that Eq.(\ref{lrckpz}) is 
invariant under $h\rightarrow -h$ and $\lambda \rightarrow -\lambda$ transform.
Therefore we consider both positive and negative values of 
$U_{\rho}$ and take $U_{0}\geq 0$  without any loss in generality.
As shown in FIG.~1, there are various  
($U_{\rho}$, $U_{0}$) phase diagrams depending on the 
dimensionality $d$ and the long-range interaction parameter $\rho$.
We explain the various phase diagrams in detail.

I) $\rho>0$: The effective nonlinearity $U_{\rho}$ is dominant than
$U_{0}$, then the phase in all space ($U_{\rho},U_{0}$) except
for $U_{\rho}=0$ is determined by the long-range $\lambda_{\rho}$ term in 
Eq.~(\ref{lrckpz}). 
For $d<2-\rho$, the LR fixed point is stable, and the surface is the LR 
rough phase with the positive roughness exponent 
(we call it the LR rough) given by Eq.~(\ref{exponent}).
If $U_{\rho}=0$, the SR rough phase with the positive roughness
exponent($\chi=(2-d)/3$, we call it the SR rough) 
exists such that a phase transition takes place 
between two LR rough phases when the sign of $U_{\rho}$ is changed. 
The critical behavior($U_{\rho}=0$) follows the SGG's nonlinear equation 
with the SR rough phase.
For $d=2-\rho$, the surface is the logarithmically rough phase
with a zero roughness exponent (we call it the Log rough).
For $2-\rho < d<2+2\rho$, the phase is controlled by the LR fixed point
and the
\begin{figure}[htb]
\centerline{
        \epsfxsize=90mm       
\centerline{ \epsffile{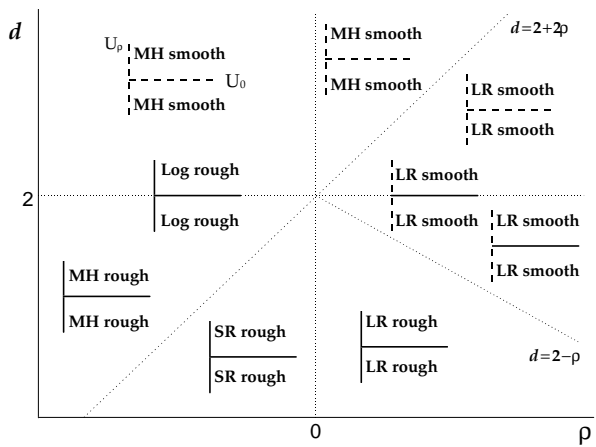}  }
        \vspace*{0.1cm}
        }
\caption {
$U_{\rho}$($y$-axis) vs $U_{0}$($x$-axis) phase diagram in ($\rho$, $d$)
space. On the $x$-axis and $y$-axis of the phase diagrams, 
the solid lines denote the rough phase and
the dotted lines do a smooth phase.
The detailed meanings of the corresponding rough or smooth phase
are explained at Table~1.
For $\rho>0$, the long-range effect makes the surface less rough than a 
case of $\rho=0$ (see, the region $2-\rho \leq d \leq 2$ for $\rho>0$). }
\label{flow}
\end{figure}
\bigskip
$\!\!\!\!\!\!$the surface is the LR smooth phase 
due to the negative value of the 
roughness exponent ($\chi=(2-d-\rho)/3$, we call it the LR smooth). 
In addition, the various critical behaviors depending on
both value of $\rho$ and dimension $d$ are shown in FIG.~1.
Phases at the critical line($U_{\rho}=0$) are
SR rough for $2-\rho \leq d<2$, Log rough for $d=2$,
and MH smooth for $2<d<2+2\rho$.
Here, the MH smooth phase is defined by the linear equation with a 
negative roughness exponent ($\chi=(2-d)/2$, we call it the 
Mullins-Herring(MH) smooth).
For $d \geq 2+2\rho$, both the LR and SR fixed points are irrelevant, so that
only the MH smooth phase of the linear equation exists.
Therefore, various phase transitions take place when the sign of $U_{\rho}$
is changed, except for the region ($d \geq 2+2\rho$) where no phase
transition occurs for all values of $U_{\rho}$ and $U_{0}$.
At physical dimension $d=2$, it is well known that, 
for the short-range interaction($U_{\rho}=0$ and $U_{0}\neq 0$), 
the SR fixed point(SGG) is irrelevant, so that the surface is  
logarithmically rough.
However, if $U_{\rho} \neq 0$ and $\rho>0$ (that is, for the long-range
interaction), the LR fixed point is 
relevant and the surface becomes LR smooth with the negative exponents given 
by Eq.(\ref{exponent}).
We thus find that nonzero $U_{\rho}$ term with $\rho>0$ can make the surface
less rough than the logarithmically rough of the 
case $\rho=0$ (see Table~1).

II) $\rho<0$ : The LR fixed point is irrelevant on the ground that
$U_{0}$ is dominant than $U_{\rho}$.
So, the short-range term in Eq.~(\ref{lrckpz}) corresponding to the
SGG equation determines the surface behavior in all space
($U_{\rho},U_{0}$) except for $U_{0}=0$.
For $d<2+2\rho$, the SR fixed point is stable and the surface is always
SR rough except for $U_{0}=0$, when it is a LR rough phase. 
So, there is no phase transition for $U_{0} \neq 0$. 
For $d \geq 2+2\rho$, both the SR and LR fixed points are no longer
stable, so that the phase is 

\begin{center}
{\footnotesize
\begin{tabular}{|c|c|c|c|} \hline \hline
    &   &Phase of $U_{0}(x{\mbox -}axis)$ & Phase of 
                               $U_{\rho}(y{\mbox -}axis)$  \\ \hline
    & $d<2+2\rho$        & SR rough & LR rough \\ \cline{2-4}
$\rho<0$ & $2+2\rho \leq d<2$ & MH rough & MH rough \\ \cline{2-4}
    & $d=2$   & Log rough & Log rough \\ \cline{2-4}
    & $d>2$   & MH smooth & MH smooth \\ \hline \hline
    & $d<2-\rho$ & SR rough  & LR rough  \\ \cline{2-4}
    & $d=2-\rho$ & SR rough  & Log rough  \\ \cline{2-4}
    & $2-\rho < d <2$  & SR rough &      \\ \cline{2-3}
$\rho>0$ & $d=2$ & Log rough & LR smooth \\ \cline{2-3}
    & $2<d<2+2\rho$ & MH smooth &     \\ \cline{2-4}
    & $d \geq 2+2\rho$ & MH smooth & MH smooth \\ \hline \hline
\end{tabular}
}
\end{center}

{\small
Table~1. Various phases depend on both $\rho$ and $d$. 
These phases correspond to the diagrams in FIG.~1.
There are six different phases:
LR rough if $\chi=(2-d-\rho)/3$ is positive, 
LR smooth if $\chi=(2-d-\rho)/3$ is negative,
SR rough if $\chi=(2-d)/3$ is positive, 
MH rough if $\chi=(2-d)/2$ is positive, 
MH smooth if $\chi=(2-d)/2$ is negative, 
and Log rough if $\chi=0$.
}
\bigskip
\newline
\newline
$\!\!\!\!\!$governed by the linear equation.
Therefore, the phases are MH rough for $2+2\rho \leq d<2$, logarithmically
rough for $d=2$, and MH smooth for $d>2$.   
Here, the MH rough phase means the positive roughness exponent
given by the linear equation ($\chi=(2-d)/2$).
Unlike the case of $\rho >0$, any phase transitions do not take 
place for all spaces of $U_{\rho}$ and $U_{0}$.

We have also studied Eq.(\ref{lrckpz}) with a nonconservative noise $\eta$ 
instead of a conservative noise $\eta_{c}$.
The nonconservative noise $\eta$ is a white noise of zero mean with
$\langle \eta({\bf r},t) \eta({\bf r}',t')\rangle
=2D \delta({\bf r}-{\bf r}')\delta(t-t')$. 
There are two sets of axial fixed points. The phase diagrams in this 
growing surfaces are essentially the same as the FIG.~1
if the dimensionality is replaced by $d\rightarrow d-2$.
At the SR fixed point,
$d_{c}=4$, $\chi =(4-d)/3$ and $z=(8+d)/3$, in agreement with those of
the equation introduced by Lai and Das Sarma\cite{laidas}.
At the new LR fixed point with $d_{c}=4+2\rho$, the exponents are given
by $\chi =(4-d-\rho)/3$ and $z=(8+d-2\rho)/3$ which are obtained from
the non-renormalization of $D$ and $\lambda_{\rho}$ 
in one-loop approximation. 
Experimental results for the growth of Fe films on Fe(001) using 
high-resolution
low-energy electron diffraction(HRLEED) technique\cite{he} show 
$\chi =0.79\pm 0.05$ and $\beta=\chi /z =0.22\pm 0.02$ for $d=2$. 
These values are consistent with our results with the value $\rho=-0.37$. 
At this point, it is unclear whether this experimental system really possesses
the long-range interaction as that of Eq.(\ref{lrckpz}). 
It is thus strongly invited to
examine other systems with the long-range interactions.

In summary, we have studied the conserved KPZ equation in the presence
of long-range interactions. For positive value of $\rho$, the
long-range nonlinear term makes the surface less rough and 
produces the different values of the exponents compared to them
of the SGG equation.
Particularly, at physical dimension $d=2$, the surface is 
smooth for $\rho >0$, 
while the surface is logarithmically rough for $\rho =0$.
However, the long-range nonlinear term becomes irrelevant for the negative
value of $\rho$, and the surface is controlled by the SGG fixed points.

\bigskip
This work is supported in part by
the Ministry of Education (No. BSRI-97-2409),
and by the Korea Science and Engineering Foundation through 
the Scientific Research Center of Excellence program of the Seoul National
University Center for Theoretical Physics.


\begin{thebibliography}{99}
\bibitem{janssen} However, following H. K. Janssen,
       Phys. Rev. Lett. {\bf 78}, 1082 (1997),
the relation $z+\chi =4$ should be modified as $z+\chi =4-3\delta$. 
By the two-loop RG calculation, it has been shown that $\delta$ is
very small. Then, even if there exists such a correction, the exponents
$\chi$ and $z$ may not deviate seriously from the results of the one-loop 
approximation. From this result of two-loop approximation, 
because $\lambda_{\rho}$ may not be renormalized,
it is expected for $z+\chi =4-\rho$ to be affected 
only by the small quantity $\delta$.
However, in this paper we will use the results of one-loop approximation.
\end{thebibliography}
\end{document}